\documentclass[preprint]{aastex}
\usepackage{graphicx}
\usepackage{amssymb}
\def\cxo{{\em CXO\ }}

\begin{document}

\title{The helical jet of the Vela Pulsar}
\author{Martin Durant$^{1,2}$, Oleg Kargaltsev$^{1,3}$, George G. Pavlov$^{4,5}$, Julia Kropotina$^{5,6}$, Kseniya Levenfish$^{5,6}$}
\affil{$^1$ University of Florida, 211 Bryant Space Science Center, Gainesville, FL, USA\\
$^2$ Department of Medical Biophysics, University of Toronto, M4N 3M5 Canada\\
$^3$ The George Washington University, Department of Physics, Washington, DC 20052, USA\\
$^4$ Pennsylvania State University, 525 Davey Lab, University Park, PA, USA\\
$^5$ St.-Petersburg State Polytechnical University, Polytekhnicheskaya ul.\ 29, St.-Petersburg,
195251, Russia\\
$^6$ Ioffe Physical-Technical Institute, St.-Petersburg, Russia\\}
\email{mdurant@sri.utoronto.ca}

\begin{abstract}
We have studied  the fascinating dynamics of the nearby Vela pulsar's nebula in a campaign comprising eleven  40\,ks observations with Chandra X-ray Observatory ({\sl CXO}). The  deepest yet images revealed the shape, structure, and motion of the 2-arcminute-long pulsar jet. We find that  the jet's shape and dynamics are remarkably consistent with    that of a steadily  turning  helix projected on the sky. We discuss possible implications of our results, including free precession of the neutron star and MHD instability scenarios.
\end{abstract}
\maketitle

\section{Introduction}
The rapid rotation of a pulsar's magnetosphere gives rise to strong electric fields, particle acceleration and  pulsed radiation across the whole EM spectrum. Pulsar wind and associated radiation can fill a substantial volume in the vicinity of a pulsar and be observable throughout the EM spectrum as a pulsar wind nebula (PWN; for recent reviews of PWNe  see \citealt{2006ARA&A..44...17G,2008AIPC..983..171K}). The termination shock, where bulk flow kinetic energy is converted into internal energy of the plasma, is resolved for the case of the Crab Nebula (and some other PWNe) as the inner edge of the torus seen in X-rays \citep{2000ApJ...536L..81W,2008AIPC..983..171K}.
 In addition to toroidal structures, many PWNe exhibit jets which are believed to be outflows along the pulsar's spin axis \citep{2008AIPC..983..171K}. The formation, confinement and structure of jets  are still poorly understood, particularly  for pulsar jets, which are very different from  AGN, X-ray binary and protostellar jets in some respects (see, e.g., \citealt{2008JPhCS.131a2052H}	
 for a review).  The proximity of the Vela PWN offers a unique opportunity to carry out a detailed, time-resolved  study of  jet dynamics on timescales of months. Such studies can shed  new light on  jet physics in general, but they  are impossible to perform for AGNs jets, which are the most commonly discussed examples of magnetized MHD outflows.  

The Vela pulsar is a young (pulsation period $P=89$\,ms and spin-down age $\tau=11.3$\,kyr), energetic pulsar  located  at a distance $d\approx300$\,pc. The pulsar is famous for its $\gamma$-ray brightness \citep{1999ApJS..123...79H} and glitching behavior \citep{1990Natur.346..822M}. The latter was considered strong evidence of superfluidity in the pulsar core \citep{1999MNRAS.307..365S}.  The Vela pulsar powers the nearest bright X-ray PWN, with a puzzling double-arc structure \citep{2001ApJ...556..380H,2001ApJ...552L.129P,2003ApJ...591.1157P}. 

Previous X-ray observations with the {\sl Chandra X-ray Observatory} (\cxo)
 led to the discovery of the pulsar's outer jet (beyond the arcs, in the direction of the proper motion), which is faint compared to the inner nebula, and has a peculiar curved morphology \citep{2001ApJ...554L.189P,2003ApJ...591.1157P}. Unfortunately, the read-out streak, the chip gaps,  and the suboptimal cadence hampered modeling of the outflow dynamics. Here we present more recent \cxo follow-up observations (3 images in 2009 and 8 in 2010, a total of $\approx$440\,ks of new data), with each image twice as deep as in the previous observations. 
 In this Letter we report the first results of these observations.

\section{Observations and Data reduction}
\begin{figure*}
\begin{center}
\includegraphics[width=0.95\hsize]{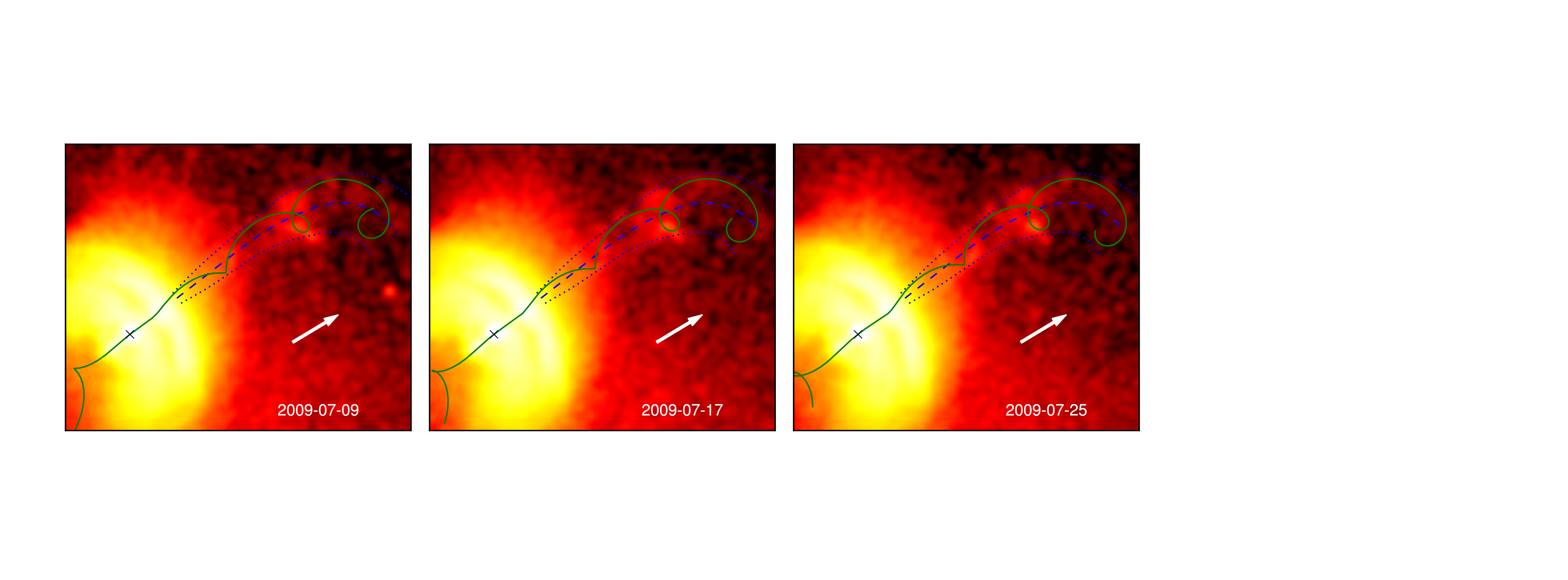}
\caption{The 2009 data with the projected trace of the best-fit helical model superimposed (green). The envelope and guiding line (see text) are also marked (dotted and dashed blue lines, respectively). A black X marks the location of the pulsar and a white vector the direction of proper motion on each image. The parameters for the helix are $\tau=122$\,day, $\lambda = 0.23$\,ly and an opening half-angle of 5\degr.}\label{pics}
\end{center}
\end{figure*}

\begin{figure*}
\begin{center}
\includegraphics[width=0.9\hsize]{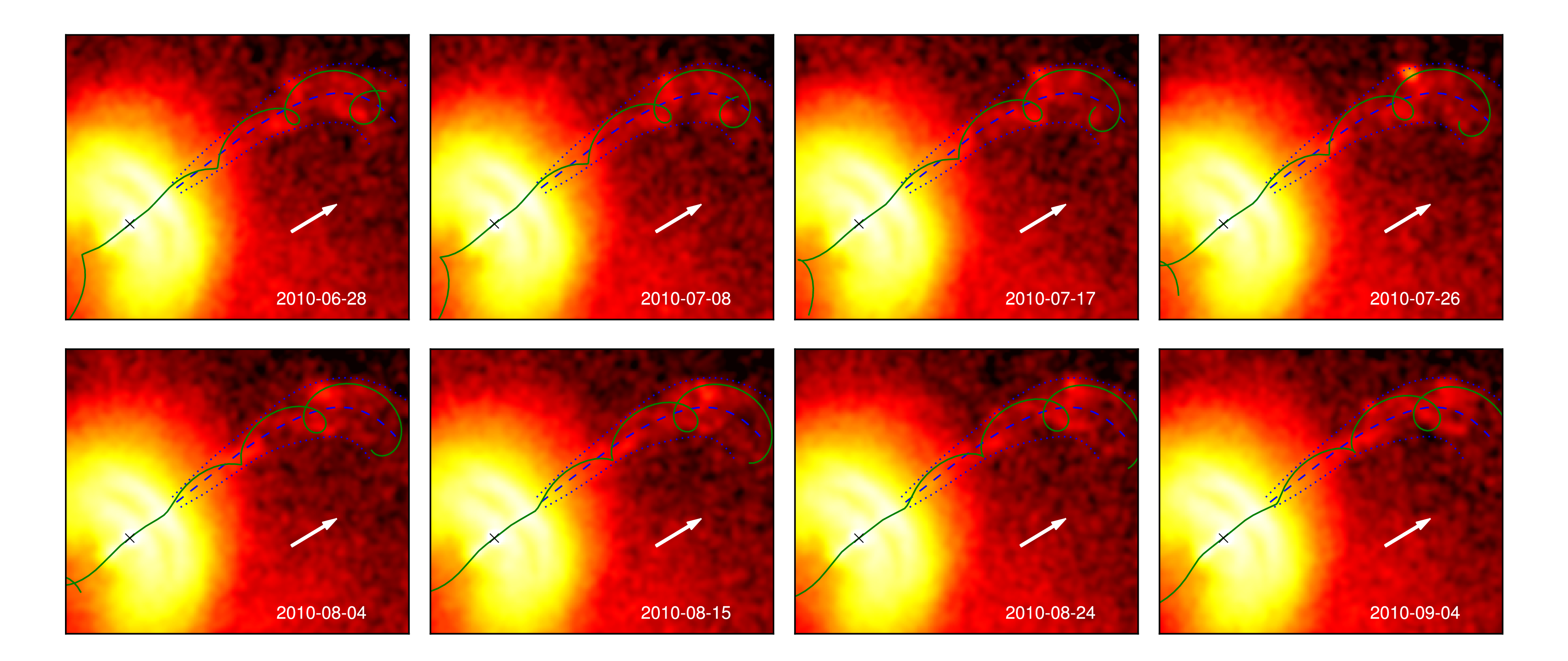}
\caption{As Figure \ref{pics}, but for the 2010 data.}\label{pics2}
\end{center}
\end{figure*}

We observed the Vela Nebula with the Advanced CCD Imaging Spectrometer (ACIS) aboard {\sl Chandra}. Three observations were performed in July 2009 and eight in June-September 2010. Each observation had an exposure time of about 40\,ks, and the intervals between successive observations were about a week. The FAINT telemetry mode, with a frame time of 3.24\,s, was used
with the target 
imaged on the back-illuminated S3 chip. The read-out streak (caused by the bright pulsar) did not overlap with the jet. We applied standard filtering to the recorded events and ignored energies outside the 0.5--8\,keV band. The sequence of the resultant images is shown in Figure \ref{pics}-\ref{pics2}.

\section{Analysis and Model}

The jet in the direction of the pulsar proper motion (northwest) has two components: the bright {\sl inner} jet, which extends for $\approx13''$ from the pulsar up to the outer arc of the PWN, and the {\sl outer} jet, which is much fainter but extends further for another $\approx100''$.
 For brevity, hereafter we will refer to the outer northwest (NW) jet as ``the jet'' unless it causes ambiguity.
  With new deeper images (compared to those in \citealt{2003ApJ...591.1157P}), the fainter emission filling the jet is clearly seen  in addition to the brighter knots
   (see Figures~\ref{pics} and \ref{pics2}).
Despite the changing shape, the 
 jet as a whole remains confined to a fairly well defined region (jet ``envelope''),
 which is initially narrow and pointing in the direction of the pulsar's proper motion (as does the inner jet), 
  but then gradually bends towards the southwest (SW) at greater distances from the pulsar (see Figures 1 and 2, blue curves; also below). 

The jet shape and dynamics (see Figures 1 and 2 and the corresponding animation)
 strikingly resemble that of a rotating helix projected on the sky.  The successive positions of the  knots detected within the jet in several images, do not appear to have a large velocity component transverse to the jet but rather  move  approximately radially away from the pulsar. The brighter inner  jet does not show nearly as much motion as the outer jet. However, the point at which the outer jet emerges beyond the outer arc (see \citealt{2001ApJ...552L.129P} for definitions)  changes position with time, moving back and forth along the outer arc edge. This remarkable phenomenology prompted us to explore a simple geometrical model where a small perturbation, which initially occurs close to the base of the jet, propagates ballistically and grows with the distance from the pulsar 
resulting in  a three-dimensional helix, which we view projected onto the sky. 

In our model we adopt the geometric orientation suggested by \citet{2001ApJ...556..380H},  which implies that the angle between the axis of the inner jet (the latter is directed along the pulsar's spin axis  and  the proper motion direction) and the line-of-sight (LoS, normal to the sky) is 53\degr, with the inner jet pointed away from the observer (i.e., behind the plane of the sky). The ballistic 
  motion of the material in the (outer) jet is 
  gradually modified 
    due to the global bending of the jet (discussed by \citealt{2003ApJ...591.1157P}; see also the dashed blue curve in Figures~1 and 2). We require the helix to be confined within a bent envelope (see blue curves in Figures~1 and 2), which we define at the outer boundary of the jet emission on the sky,  from the sum of all 11 images. We define the ``guiding'' line as a curve going through the middle of the envelope. The helix radius increases with distance from the pulsar (forming a cone with opening half-angle 5$^\circ\pm1^\circ$, as measured from the pulsar). The helix and its  trace (projection) on the sky were constructed in the following way.  The helix was defined parametrically using a rotating vector and a moving circle.
    At each point along the guiding line (dashed blue curve in Figures 1 and 2) we defined a circle in a plane locally orthogonal to the guiding line. The circle  radii are increasing with distance from the pulsar, so that each of  the circles meets the jet envelope (dotted curves in Figures 1 and 2). For each circle we drew a vector from the circle's center to a point on the circle. We then rotated these vectors  in their circle planes so that the azimuthal angle (in the circle plane) changes  linearly  with the distance from the pulsar.
   The endpoints of these  vectors trace  a 3-dimensional helix, which we 
  projected onto the sky.  The dynamics was introduced by having the azimuthal  angle of the first vector (near the pulsar) change linearly with time, completing one turn in time $\tau$.     Thus, by construction,  in our model the speed of a helical pattern  is  constant, $v=\lambda/\tau$, where  $\lambda$ is the helix pitch (the distance between successive turns of the helix).
      


As a next step, we  fitted our dynamical model to the \cxo image series by varying $\tau$, $\lambda$, and the  initial azimuthal  angle of the vector near the pulsar
(which sets the jet ``launching'' direction at a given time).  
The model parameters were optimized to maximize  the number of photons falling in a 3\,pix-wide stripe along the trace. The best match to the data corresponds to the stripe containing $\approx500$ extra counts ($\approx$4200 counts total) compared to the average value for randomly chosen parameters ($\approx3700$ counts total). 
To obtain the most conservative estimate of this significance we performed extensive  Monte-Carlo simulations. In particular we simulated the jet shapes for 50$\leq\tau\leq$200\,d and 0.1$\leq\lambda\leq$0.3\,ly and any initial azimuthal angle, which are also the ranges we used in our best-fit search. The simulations resulted in 10000 realizations for the jet shape. We then calculated the ratio of the number of realizations where the jet's trace intercepts $>$4200 photons to the total number of MC realizations. This resulted in  the chance coincidence probability of 0.4\% which corresponds to $\approx3\sigma$ significance.  A simplistic estimate based on Poisson counting statistics leads to larger significance. 
Because the data consist of two series with about a one-year gap in-between, there is strong aliasing in the goodness-of-fit measure as a function of $\tau$, resulting in several acceptable periods, $\tau=73\pm2, 91\pm5, 129\pm10$\,d. We repeated this  analysis including older observations from 2000-2002. The Monte-Carlo significance did not improve, but one of the above solutions was preferred above the others:  $\lambda=0.230\pm0.014$\,ly (corresponding to a projected angular distance of 40$^{\prime\prime}$ on the sky) and $\tau=122\pm5$\,d ($\approx$3\,cycles/year), implying $v=\lambda/\tau = 0.23c$\,yr$/0.33$\,yr = $0.7c$. This flow speed is in agreement with the earlier estimates of the apparent speed of individual knots  in the jet, $v=0.3c-0.7c$ \citep{2003ApJ...591.1157P}. 
We have sketched the helical trace corresponding to the best fit on the images in Figures \ref{pics} and \ref{pics2} (green curves).

%

\section{Discussion}

The  model described in Section 3
appears to provide a simple description of the observed jet 
 geometry and dynamics. According to our model,  the flow of particles  is {\sl ballistic}: the bulk velocity is predominantly radial (away from the pulsar)  in the direction  defined by its launching angle, and the helix arises only from the steady change of that launching direction (similar to a rotary garden sprinkler). Indeed, the individual knots within the jet appear to move away from the pulsar radially, rather than following the
helical shape of the jet (Figures 1 and 2).
 Below we discuss the implications of a helical structure produced by a ballistic flow.
 

Helical patterns are  often seen in  AGN jets (e.g., \citealt{2012arXiv1202.1182P}), protostars (e.g., \citealt{2007ApJ...670.1188L}) and accreting binary systems (SS443; \citealt{2002MNRAS.337..657S}). Among the possible mechanisms leading to a helical structure, the most frequently mentioned are the helical instability mode  in a magnetized plasma column (the `kink' instability; \citealt{2011IAUS..275...41H,2006AIPC..856...57H}  and references therein) or  jet precession  \citep{2006ApJ...645...83V,2002ApJ...580..950M}.  Perhaps the clearest example of a Galactic helical jet  is that of the micro-quasar SS 433, where the shape of the jet has been explained as the motion of a matter stream ejected from a precessing nozzle located in the vicinity of the compact object  \citep{2002MNRAS.337..657S}. Precession in this latter case is torque-driven through the interaction of the compact object with its binary companion.   Also in AGN jet models involving precession  \citep{1999A&A...344...61A,2004ApJ...608..149T}, the precession  is torque-driven via interactions between the accretor and the accretion disk. An alternative magneto-hydrodynamic (MHD) instability scenario \citep{2011IAUS..275...41H,2006AIPC..856...57H} does not require any external force: rather, the helical pattern grows from a small perturbation which may be any random fluctuation.

For the jet of the  Vela pulsar (which lacks any evidence of accretion or a companion), we consider two mechanisms which could give rise to helical structure: free precession (i.e., without external torque), and a kink MHD instability in the jet flow. Whilst the first is conceptually simple, requiring only oblateness of the pulsar, its existence  has not yet been   established  in any pulsar (although several claims have been made, see below).
 The latter option has undergone extensive theoretical and numerical investigations for AGN jets, but not pulsar jets, whose properties are different. Moreover, the two mechanisms need not be mutually exclusive because the initial small perturbation in the jet flow can be caused by precession and then be amplified via kink instability.

The mechanism responsible for pulsar jet formation is a matter of debate.
If the jets are formed at a relatively large distances from a pulsar, e.g., via a magnetically channeled backflow of particles from the equatorial post-shock region\footnote{A possibility suggested by \citet{2002MNRAS.336L..53B,2003AstL...29..495K} and reproduced in the numerical simulations of \citet{2004MNRAS.349..779K,2004A&A...421.1063D,2006A&A...453..621D}.}, then one might expect the effects of a free precession on a jet be somewhat washed out. On the other hand, it is possible that despite the presence of the backflow at larger distances, the jet is actually formed much closer to the pulsar \citep{2012arXiv1209.1589L} and then the impact of  precession on the jet's behavior would be more pronounced.

\subsection{Helical structure due to precession}

The Vela pulsar spin axis may be freely precessing with the period we measured for the jet from our model, in which case the jet is simply launched ballistically along the spin axis. Previous hints of free precession in pulsars have been reported, based on (quasi-)periodic changes in pulsar spectra, pulse profiles or timing residuals \citep{2006A&A...451L..17H,2011arXiv1107.3503J}. The periodic signature seen in PSR B1828$-$11  was interpreted as precession \citep{2000Natur.406..484S}, but after ten years of further observation, this interpretation has not been confirmed. The varying X-ray flux of the isolated neutron star RX J0720.4$-$3125 was  suggested to be due to precession but, again, further observations have failed to firmly establish this conclusion \citep{2009A&A...498..811H,2012MNRAS.419.1525H};
the periodic flux variations could plausibly be caused by effects other than precession \citep{2012arXiv1202.1123Z}. \citet{2008Ap&SS.317..175P}, commenting on the putative precession of PSR B1828$-$11, argued for a relationship between the precession period $\tau$ and the pulsar spin period $P$ as $\tau\sim kP^{1/2}$, which would predict $\tau=120$\,d for Vela.  Similarly, \citet{2011arXiv1107.3503J} measured the periodic modulations of several pulsars, finding a correlation between the long-term period and the pulse period, which could be ascribed to precession.  Periods in the 80--200\,d range are predicted for a pulsar with $P$=89\,ms (see Figure 1 in  \citealt{2011arXiv1107.3503J}). Therefore, it is conceivable that the helical shape of the Vela pulsar  jet could be due to  free precession.  If the Vela pulsar is indeed precessing,  long-term timing in  radio may be able to detect it \citep{1996ASPC..105..101D}, but such detections are complicated by glitching. If precession is confirmed for the Vela pulsar, it  should be present  in other pulsars as well. Interestingly, recent analysis by \citet{2011evhe.confE...2W}  shows that the Crab pulsar jet also has a corkscrew structure which evolves with time.

The precession of a pulsar has interesting theoretical implications.  For our helical model,  we can infer the stellar oblateness parameter $\epsilon=(I_3-I_1)/I_1 = P_{\rm spin}/P_{\rm prec}\approx8\times10^{-9}$ (where $I_1$ and $I_3$ are principal moments of inertia). As pointed out by \citet{2001MNRAS.324..811J}, this imposes a limit on how much of the pulsar mass can be in the superfluid interior. One might expect the superfuidity to damp any precession effects, but the coupling between crust and core is complex, and the core may partially participate in the precession \citep{2004ApJ...613.1157L}. Indeed, \citet{2008Ap&SS.317..175P}  suggests that glitching --  exchange of angular momentum between the crust and the core -- can actually be the excitation mechanism for precession. If so, then it is not surprising that Vela, which is such a prolific glitcher, should show the clearest signs of precession.

\begin{figure}
\begin{center}
\includegraphics[width=0.95\hsize]{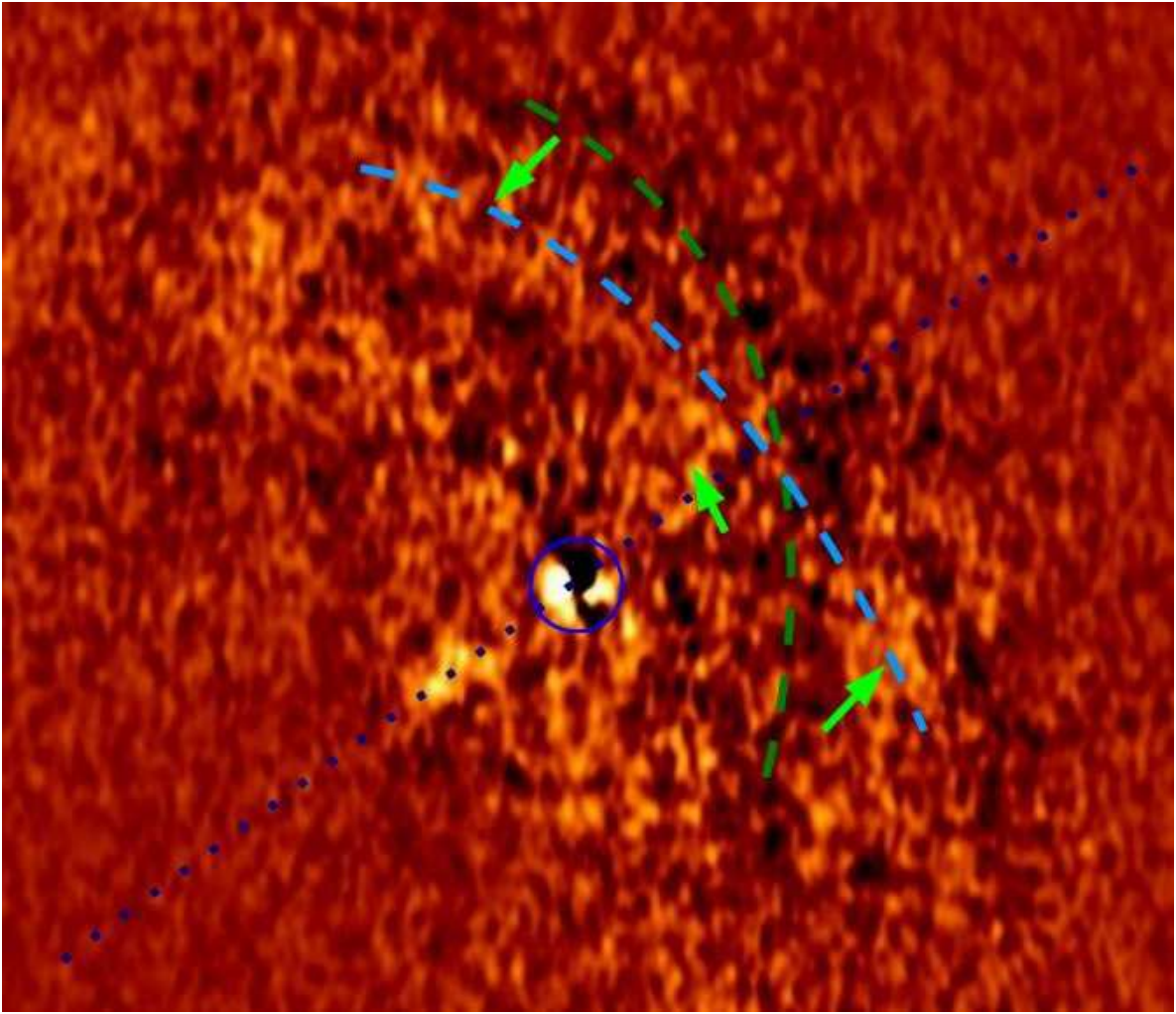}
\caption{The difference image between the first and last observation of the 2010 series. We label the axis of the inner jets (blue dotted line), the pulsar position (blue circle), and the arc positions in the first image (darker region with green dashed line) and last image (brighter region with cyan dashed line). Green arrows show the direction of apparent motion, in both the arcs and the jet. 
}\label{diffs}
\end{center}
\end{figure}

We would also like to point out that the rest of the bright inner PWN also appears to be variable on a similar timescale.   For example, Figure \ref{diffs} shows the difference image throughout the 2010 data (last image of the series minus the first image), revealing the changes in the inner arc-like structures and the inner south-eastern jet. Changes, labelled in Figure \ref{diffs}, can be easily seen. We defer detailed investigation of the geometric and spectral evolution of the inner nebula for a separate paper. It should be noted, however, that the only mechanism capable of producing a coordinated (coherent) change in the arcs (likely brightened limbs of the   equatorial outflow; \citealt{2003ApJ...591.1157P}) and both jets must be acting at the origin of these outflows (i.e. at the pulsar) making the precession  a natural candidate.   
Changes in the Vela PWN morphology seen in earlier {\sl CXO} images were suggested to be caused
by precession \citep{2007ApJ...656.1038D}, perhaps related to  earlier (quasi-)periodic timing signatures  \citep{1996ASPC..105..101D}, suggesting a precession period of 330\,d.   We note that the Vela PWN model of \citet{2007ApJ...656.1038D} assumes  that  the arcs are  ``traces of the particle beams from the two magnetic poles at the shock front''  and  the jet is a ``manifestation of a physical flow of particles along the magnetic axis of the star, not
along the rotation axis.''  Although their assumptions clearly differ from those in our model (where the jet is along the pulsar's spin axis),  the observational implications of the free precession scenario could
be similar (regardless of the specific model for the PWN features), such as correlated periodic variability in the PWN structure.


\subsection{Helical structure due to kink instability}

 Another possible scenario for the observed jet shape and dynamics is  the kink instability. 
Although  our helical model  has  strict periodicity built-in, 
 the model may still approximate, to some extent, the behavior expected due to the kink depending on the nature of the initial perturbation (which can also be the precession). 
  The MHD helical kink instability is often invoked to interpret the AGN jet shapes (e.g., \citealt{2011IAUS..275...41H,2006AIPC..856...57H} and references therein). Numerical simulations of the kink instability in the non-linear regime can produce structures resembling a helix, with  current driven, velocity shear driven, or jet precession mechanisms being considered as viable options  for driving the kink \citep{2011IAUS..275...41H}. 
Pulsar jets were discovered relatively recently, their physical properties are still poorly known, and their modeling is still in a nascent state compared to the AGN jets. 
 Although the basic MHD principles should probably be applicable to pulsar jets,  existing AGN jet models may not be  directly applicable, because the properties of these  systems (e.g., baryon loading, bulk Lorentz factor, magnetic-to-kinetic energy ratio, jet formation mechanism, jet confinement mechanism) are likely quite different. Even in the linear approximation the analysis for the local instability modes in pulsar jets  critically depends on the poorly known magnetization parameter \citep{1998ApJ...493..291B}.  
Still, some results from the numerical modeling of AGN jets might be useful for qualitative understanding of pulsar jets.
For the MHD instability, we can crudely associate the period in our helical model, $\tau$, with the characteristic timescale of the appearance of successive kinks (which will not be truly periodic unless triggered by precession).  According to the recent MHD simulations done for AGN jets by \citet{2009ApJ...700..684M,2012arXiv1207.4949M}, the instability growth timescale for the global kink development (i.e., when the jet  resembles a spiral with a radius comparable to  the jet radius) can be crudely estimated as $t_i\sim10R_{\rm jet}/v_{\rm A}$, where $R_{\rm jet}$ is the jet radius and  $v_{\rm A}$ is the Alfv\'en speed.  Using the measured jet radius $R_{\rm jet}\approx5''$ and the  Alfv\'en speed estimated for equipartition, $v_{\rm A}=0.77c$ \citep{2003ApJ...591.1157P}, we find the instability growth timescale to be $\sim100$\,ds. 
 In the images shown in Figures 1 and 2 the distinct helical pattern typically emerges by a distance  $d_i\approx1'$ from the pulsar, which implies the flow velocity $d_i/t_i\sim0.9c$ assuming that the instability starts at much smaller distance from the pulsar than $d_i$.  Although the implied flow speed is somewhat higher than the flow speed inferred above and the apparent speeds of the knots (Pavlov et al. 2003), these are consistent given the uncertainties involved, which support the above-described scenario, if the jet's magnetic field strength is near equipartiton.
 
We note, however, that the jet retained approximately the same helical morphology between the 2010 and 2011 series which corresponds to over 400\,d. On this substantially larger timescale,  one could expect to see highly nonlinear instability development, leading to the jet disruption.
However, with the data available  we cannot tell whether the jet shape remained stable from 2009 to 2010 or if it had disrupted and regrown. We also note that the extreme bending was observed only once, in the 2000 February image (see \citealt{2003ApJ...591.1157P}), and it might have been due to a disruptive, highly-developed kink. The rareness of such events hints at a substantially larger instability timescale than the estimate above. In this case the origin of the stable helical  pattern may be better explained by pure precession. On the other hand, there may be factors leading to the delay in the nonlinear instability growth, such as the pressure of the ambient medium becoming important far enough from the pulsar. At large distances from the pulsar the jet behavior is also affected by the ram pressure due to the combined action of the pulsar's motion and a strong wind within the SNR, possibly caused by the passage of the reverse shock (Pavlov et al., 2003; \citealt{2011ApJ...740L..26C}). This ram pressure can explain the clockwise bending of the jet's end and the distortion  of the helical structure at large distances from the pulsar.  
 

\section{Conclusion}
%

The Vela pulsar system continues to be a fascinating testbed for magnetized outflow models. It may also become the
first NS known to precess freely, if our results are confirmed by further modeling of the inner PWN, by additional
X-ray observations, and/or  by  careful pulsar timing analysis, which we strongly encourage. If the precession is confirmed,
the Vela pulsar may become one of the best persistent sources of gravitational waves for the next generation of detectors, given that current detecting technology is most sensitive to signal frequencies $>$10Hz, and sensitivity is greatly enhanced by searching at a known frequency and position \citep{2011LRR....14....5P}.
 Numerical MHD simulations of pulsar jets  are also required to make further progress and to reveal the physical conditions in the jet and instability development.

\medskip\noindent{\bf Aknowledgements:}
Support for this work was provided by the National Aeronautics and Space Administration through Chandra Award  GO9-0084  issued by the Chandra X-ray Observatory Center, operated by the Smithsonian Astrophysical Observatory for the National Aeronautics Space Administration under contract NAS803060. The work was also supported by NSF grants AST-0908733 and AST-0908611.  GGP was partly supported by the Ministry  of Education and Science of the Russian Federation (Contract No. 11.G34.31.0001).
 

\end{document}